\documentclass{llncs}
\usepackage{amssymb}
\usepackage{graphicx}
\usepackage{color}
\usepackage{pstricks,pst-node,pst-tree}

\title{On the Evaluation of RDF Distribution Algorithms Implemented over Apache Spark}
\author{Olivier Cur\'e, Hubert Naacke,	Mohamed-Amine Baazizi, Bernd Amann}
\institute{Sorbonne Universit\'es, UPMC Univ Paris 06, UMR 7606, LIP6, F-75005, Paris, \\
CNRS, UMR 7606, LIP6, F-75005, Paris, France\\
\email{\{firstName.lastname\}@lip6.fr}
}

\begin{document}

\maketitle
\begin{abstract}
Querying very large RDF data sets in an efficient manner requires a sophisticated distribution strategy. 
Several innovative solutions have recently been proposed for optimizing data distribution with predefined query workloads. 
This paper presents an in-depth analysis and experimental comparison of five representative and complementary distribution approaches.
For achieving fair experimental results, we are using Apache Spark as a common parallel computing framework by rewriting the concerned algorithms using the Spark API. 
Spark provides guarantees in terms of fault tolerance, high availability and scalability which are essential in such systems. Our different implementations aim to highlight the fundamental implementation-independent characteristics of each approach in terms of data preparation, load balancing, data replication and to some extent to query answering cost and performance. The presented measures are obtained by testing each system on one synthetic and one real-world data set over query workloads with differing characteristics and different partitioning constraints.
\end{abstract}

\section{Introduction}
During the last few years, an important number of papers have been published on the distribution issue in RDF database systems, \cite{DBLP:journals/pvldb/HuangAR11}, \cite{DBLP:journals/corr/abs-1212-5636}, \cite{Wu:2014:SSD:2661829.2661876}, \cite{DBLP:conf/sigmod/GurajadaSMT14} and  \cite{DBLP:journals/pvldb/HammoudRNBS15} to name a few. 
The main motivation of this research movement is the efficient management of ever growing size of produced RDF data sets, $i.e.$, repositories of hundreds of millions to billions of RDF triples are now more and more frequent. 
Being one of the popular data model of the Big data ecosystem, RDF has to cope with issues such as scalability, high availability, fault tolerance. Systems addressing these issues, $e.g.$, with NoSQL systems \cite{Sadalage:2012:NDB:2381014}, 
generally adopt a scale-out approach consisting of distributing both data storage and processing over a cluster of commodity hardware.

Depending on the data model, it is well-known that an optimal distribution, $e.g.$, in terms of data replication rate, load balancing and query answering performance, may be hard to achieve. Each distribution approach also comes with a set of data transformation and processing steps that are more or less intensive. 

Concerning graphs in general, obtaining a balanced partitioning is known to be an NP-hard problem. Hence, most systems are proposing heuristic-based approaches which tend to produce distribution with interesting properties. In a query processing context, one of the supreme properties is the ability to limit the amount of data exchanged over the network constituting the cluster. In fact, with distributed join processing, a machine may have to transfer a large locally computed temporary result to another machine for further processing. In such situations, the total duration of the query answering process can largely be dominated by the exchange of large data chunks, $e.g.$, hundreds or thousands of Gigabytes are not uncommon, over the cluster network. 

The first systems considering distributed storage and query answering for RDF data appeared quite early in the history of RDF. Systems like Edutella \cite{DBLP:conf/www/NejdlWQDSNNPR02} and RDFPeers \cite{cai2004rdfpeers} were already tackling partitioning issues in the early 2000s. More recently, systems like YARS2 \cite{DBLP:conf/semweb/HarthUHD07} and Virtuoso \cite{DBLP:journals/debu/Erling12} were based on hashing one of the RDF triple components, most frequently the subject. In 2011, \cite{DBLP:journals/pvldb/HuangAR11} (henceforth denoted nHopDB) was the first attempt to use a graph partitioning approach to fragment an RDF dataset. This system has reinvigorated the research community on this topic. Recent systems are either extending the graph partitioning approach, $e.g.$, WARP \cite{DBLP:conf/icde/HoseS13} or are complaining about their limitations, $e.g.$, SHAPE \cite{DBLP:journals/pvldb/LeeL13}. 

As a consequence of the plethora of distribution strategies, it is not always easy to identify the most efficient solution in a given context. 
The first objective of this paper is to clarify this situation by conducting evaluations of leading RDF triple distribution algorithms. A second goal is to consider Apache Spark as the parallel computing framework for hosting these implementations. This is particularly relevant in a context where a large portion of existing RDF distributed databases, $e.g.$, nHopDB, Semstore \cite{Wu:2014:SSD:2661829.2661876}, SHAPE \cite{DBLP:journals/pvldb/LeeL13}, SHARD \cite{Rohloff:2010:HMS:1940747.1940751}, have been implemented using Apache Hadoop, $i.e.$, the open source MapReduce \cite{DBLP:conf/osdi/DeanG04} reference implementation.
In \cite{DBLP:journals/cacm/StonebrakerADMPPR10}, limitations of considering MapReduce as a database system have been identified, some of them being related to the high rate of disk reads and writes.
Spark is precisely more efficient, up to 100 times, than Hadoop because it tends to work with data stored in the main memory.

Our experimentation is conducted over a reimplementation of five approaches, two hash-based, two based on graph partitioning and an hybrid one. Each system is evaluated on two datasets, one synthetic and one real-world, 
over varying cluster settings and on a total of six queries which differ in terms of their shape, $e.g.$, star and property chains, and selectivity. 
We present and analyze  experimentations conducted in terms of the time required to prepare the data, load balancing, data replication rate and query answering performance.

\section{Background knowledge}
\subsection{RDF - SPARQL}
RDF is a schema-free data model that permits to describe data on the Web. It is usually considered as the cornerstone of the Semantic Web and the Web of Data.
Assuming disjoint infinite sets U (RDF URI references), B (blank nodes) and L (literals), a triple (s,p,o) $\in$ (U $\cup$ B) x U x (U $\cup$ B $\cup$ L) is called an RDF triple with s, p and o 
respectively being the subject, predicate and object. 
We now also assume that V is an infinite set of variables and that it is disjoint with U, B and L. We can recursively define a SPARQL\footnote{http://www.w3.org/TR/rdf-sparql-query/} triple pattern as follows: 
(i) a triple $tp \in$ (U $\cup$ V) x (U $\cup$ V) x (U $\cup$ V $\cup$ L) is a SPARQL triple pattern, (ii) if $tp_1$ and $tp_2$ are triple patterns, then ($tp_1 . tp_2$) represents a group of triple patterns 
that must all match, ($tp_1$ \texttt{OPTIONAL} $tp_2$) where $tp_2$ is a set of patterns that may extend the solution induced by $tp_1$, and ($tp_1$ \texttt{UNION} $tp_2$), denoting pattern alternatives, are 
triple patterns and (iii) if $tp$ is a triple pattern and C is a built-in condition then the expression ($tp$  \texttt{FILTER} C) is a triple pattern 
that enables to restrict the solutions of a triple pattern match according to the expression C.
The SPARQL syntax follows the select-from-where approach of SQL queries. The \texttt{SELECT} clause specifies the variables appearing in the query result set.

\subsection{Apache Spark} 
Apache Spark \cite{DBLP:conf/hotcloud/ZahariaCFSS10} is a cluster computing framework whose design and implementation started at UC Berkeley's AMPlab. 
Just like Apache Hadoop, Spark enables parallel computations on unreliable machines and automatically handles locality-aware scheduling, fault tolerance and load balancing tasks. While both systems are based on a data flow computation model, Spark is more efficient than Hadoop for applications requiring to reuse working datasets across multiple parallel operations.
This efficiency is due to Spark's Resilient Distributed Dataset (RDD) \cite{DBLP:conf/nsdi/ZahariaCDDMMFSS12}, a distributed, lineage supported fault tolerant memory abstraction that enables one to perform in-memory computations 
(when Hadoop is mainly disk-based). The Spark API also simplifies the programming tasks by integrating functions which are not natively supported in Hadoop, $e.g.,$ join, filter. 
 
\subsection{Metis graph partitioner}
Due to the complexity of partitioning a graph in an optimal manner, several methods have been defined to propose an approximation, $e.g.$, \cite{DBLP:conf/dac/FiducciaM82}. These algorithms are generally not efficient for large graphs 
where a multi-level propagation approach is frequently used, $i.e.$, the graph is coarsened until its size permits to use one of the approximate solutions, then it is uncoarsened. The Metis system \cite{Karypis:1998:FHQ:305219.305248} follows 
this approach and is known to reach its limits for graphs of about half a billion triples.
Metis takes as input an unlabeled, undirected graph and an integer value corresponding to the desired number of partitions. Its output provides a partition number for each node of the graph.
nHopDB and WARP are two recent systems that are using Metis to partition RDF graphs.

\section{Systems and distributed algorithms}
In this section, we present the main features and design principles of the distribution methods we have selected. 
We consider four different approaches which can be characterized as hash and graph partitioning based. Each category is composed of two approaches which have been used in systems and described in conference publications. 
Our fifth system corresponds to an hybrid approach that mixes a hash based approach with a replication strategy that enables to efficiently process long chain queries. Note that we do not consider systems 
that partition using a ranged-based approach since they are rarely encountered in existing systems due to their inefficiency.

\subsection{Hash based approaches}
The two approaches defined in this section correspond to families of RDF database systems rather than to specific systems (as in the next section). If not extended in a particular manner, these systems do not replicate any triples across partitions. 

\subsubsection{Random hashing:}
In a distributed random hash-based solution, the key on which the data partitioning is specified does not correspond to any particular data stored in the data model. 
For instance, the key can correspond to an internal triple identifier or to some operations over the entire triple.
The former solution is the one  adopted by the Trinity.RDF system \cite{DBLP:journals/pvldb/ZengYWSW13}. 
These two approaches do not require an additional data structure to identify the partition a particular entry is stored in. The only elements that are required for directed lookup are the hash function and the method to obtain the key.
Some other forms of random partitioning exist and may require an additional structure for directed lookups to cluster nodes where triples are located, $e.g.$ round-robin approach. We do not consider such approaches in this work since they do not guarantee nice query processing properties for any of the query shapes (star, property chains, tree, cycle or hybrid).

\subsubsection{RDF triple element hashing: }
In this system, the key provided to the hash function is one of the elements of RDF triples. The most frequent approach is to partition by triple subjects but the object or predicate element can also be considered. Partitioning by subject provides the nice property of ensuring that star-shaped queries, $i.e.$ queries composed of a graph where only one node has an out-degree greater than 1, are performed locally on a given machine. Nevertheless they do not provide guarantees for queries composed of property chains or complex query patterns. One advantage of this approach is that it does not require an additional structure to locate the partition of a given key. Systems like Yars2, Virtuoso, Jena ClusteredTDB and SHARD are adopting this approach.

\subsection{Graph partitioning based approach}\label{sec:gp}
The hash-based approaches just presented are likely to require a high data exchange rate over the network for complex query patterns, $i.e.$ those not corresponding to a star.
One way to address this issue is to either organize a replication of data and/or to analyze the query workload. 
Of course, such an organization comes at a processing cost which needs to considered with attention. 
Systems corresponding to each of these approaches are considered next. 

\subsubsection{nHopDB: }
The distribution approach presented in \cite{DBLP:journals/pvldb/HuangAR11} is composed of two steps. 
In a first stage, the RDF dataset is transformed such that it can be sent to the Metis graph partitioner, $i.e.$, remove properties and undirect the graph where subjects and objects are encoded contiguously.
Then, Metis's results are translated to triples allocation over the cluster. The partition state obtained at the end of stage 1 is denoted as 1-hop. The second stage starts and corresponds to an overlap strategy which is performed using a so-called n-hop guarantee.
Intuitively, for each partition, each leaf $l$ is extended with triples whose subject correspond to $l$. This second stage can be performed several times on the successively generated partitions.
Each execution increases the n-hop guarantee by a single unit.

\cite{DBLP:journals/pvldb/HuangAR11} describes an architecture composed of a data partitioner and a set of workers corresponding to RDF-3X \cite{DBLP:journals/vldb/NeumannW10} database instances.
Some queries can be executed locally on a single node and thus enjoy all the optimization machinery of RDF-3X. 
For queries where the answer set spans multiple partitions, the Hadoop MapReduce system is used to supervise query processing. 

\subsubsection{WARP: }
The WARP system \cite{DBLP:conf/icde/HoseS13} has been influenced by nHopDB and the Partout system \cite{DBLP:journals/corr/abs-1212-5636} (the two authors of WARP also worked on Partout). 
From the former, it borrows the graph partitioning approach and the 2-hop guarantee. Just like Partout, it then refines triple allocation by considering the query workload, $i.e.$, a set of the most frequently performed queries over this dataset. 
The system considers that this query workload is provided in one way or another. In fact, each of these queries are transformed into a set of query patterns. As a result, WARP  guarantees that some frequent queries can be processed locally without exchanging data across machines. For these queries, each partition should contain sufficient data such that the result of the query is the union of local results.
WARP proceeds as follows:
\begin{enumerate}
\item It computes a first data partitioning using the Metis graph partitioner. 
\item It fragments the data in partitions according to the subject value and loads each data partition into the independent RDF-3X \cite{DBLP:journals/vldb/NeumannW10} management system. 
\item A replication strategy is applied to ensure a 2-hop guarantee. 
\item For each query pattern, WARP computes the number of triples to replicate. To this end, it decomposes the pattern into a set of local sub-queries which are all evaluated locally. 
Each of those sub queries is a candidate to be the starting point (called seed query) for the evaluation of the entire query pattern.
The main idea of WARP  is to bring the missing triples into the partitions that contains the triples of the seed. To do so, for each candidate and partition, it computes the cost of transferring missing triples into the current partition.
Of course, it selects the seed query candidate that minimizes the cost.
\end{enumerate}


The WARP system implements its own distributed join operator to combine the local sub-queries. Locally, the queries are executed using RDF-3X machinery.

\subsection{Hybrid approach}
The design of this original hybrid approach has been motivated by our analysis of the WARP system as well as some hash-based solutions. 
We have already highlighted (to be confirmed in the next section) that the hash-based solutions require short data preparation times but come with poor query answering performances for complex query patterns. 
On the other hand, the WARP system proposes an interesting analysis of query workloads which is translated into an efficient data distribution. Next, we will see that most of data preparation for WARP is spent in the graph partitioning stage.
Hence, it seems interesting to combine a hash-based partitioning with a query workload aware refinement.





\section{Spark system implementations}

\subsection{Dataset loading and encoding}
All datasets are first loaded on the cluster's Hadoop File System(HDFS). In the experimentation section, we do not provide measures on this loading stage.
We can only stress that the loading rate in our cluster averages 520.000 triples per second. 

Like in most RDF stores, each dataset is encoded  by providing a distinct integer value to each node and edge of the graph (see \cite{cure:hal-01083133} Chapter 4 for a presentation of RDF triple encoding methods). 
The  computation is completely performed in parallel in one step using the Spark framework. We do not provide implementation details due to space limitation.\footnote{Consult http://www-bd.lip6.fr/wiki/doku.php?id=site:recherche:logiciels:rdfdist for implementation details.} The encoded datasets, together with their dictionaries (one for the properties and another for subjects and objects) are also loaded into HDFS. In all experimentations, the data is loaded within the Spark programs from HDFS. 

\subsection{Hash-based approaches}
This approach is relatively straightforward in the context of Spark which provides through its API the methods to partition a dataset.
In the case of the random-hash partitioning, the system computes the partition of a triple given its subject, predicate and object, $i.e.$, the key is the triple.
In the triple element hashing, we specify the subject as a key to the partitioning method. 
None of the implementations are extended to provide any form of replication. 
The query answering evaluation is performed forthrightly following a translation from SPARQL to Spark scripts requiring a mix of \texttt{map}, \texttt{filter}, \texttt{join} and \texttt{distinct} methods performed over RDDs.

\subsection{Graph partitioning-based approaches}
The two systems in that partitioning category require three Metis related steps: preparation, computation and transformation of the results.
Because Metis deals with unlabeled and undirected graphs, we start by removing predicates from the datasets then append the reversed subject/object pairs to the pair set yielding thus an undirected graph.
Using Metis imposes also limitations in terms of accepted graph size.
Indeed, the largest graph that can be processed  contains about half a billion nodes.
Consequently, we limit our experimentations to datasets of at most 250 million RDF triples provided that their undirected transformation yields graphs of 500 million nodes.
The output of Metis is a set of mapping assertions between a node and its partition. 
Based on these mappings, we allocate a triple to the partition of its subject.
In terms of data encoding, we extend triples with partition identifiers yielding   quads. Note that at this stage, the partition identifier can be considered as 'logical' and not 'physical' since the data is not yet stored on a given machine. 
We would like to stress that the preparation and transformation phases described above are performed in parallel using Spark programs.

Concerning the nHopDB system, the n-hop guarantee is computed over the RDD corresponding to generated quads. This Spark program can be executed (n-1) times to obtain an n-hop guarantee. 

Intuitively, our WARP implementation analyzes the query workload generalization using Spark built-in operators. 
For instance, consider the following Basic Graph Pattern (henceforth BGP) of a query denoted Q1: \texttt{?x advisor ?y . ?y worksFor ?z . ?z subOrganisation ?t}, the system uses the \texttt{filter} operator to select the triples 
that match the \texttt{advisor}, \texttt{worksFor} and \texttt{subOrganization} properties. 
Moreover, the \texttt{join} operator is used to perform join equality predicates on variables \texttt{y} and \texttt{z}. 
A query result is thus a set of bindings. We extend the notion of variable bindings with the information regarding the partition identifier of each triple.
For instance, an extract of a Q1's result (in an unencoded readable form) is represented as \texttt{\{(“Bob”,”Alice”,1), (“Alice”, “DBteam”,3),(“DBteam”, “Univ1”,1)\}}.

We see on the extracted result that the triple binding for \texttt{?y worksFor ?z} is \texttt{\{(“Alice”, “DBteam”,3)\}} (respectively \texttt{"Alice"} and \texttt{"DBTeam"} are bound to variables \texttt{?x} and \texttt{?y}) and is located on partition 3 whereas the 2 other triples are on partition 1.  
Thus we can efficiently access it to count the number of triples to replicate. For instance, if we consider the seed \texttt{(?x advisor ?y)}, we need to replicate the triple \texttt{(“Alice”, worksFor, “DBteam”)} in partition 1 
by copying it from partition 3. 
As specified earlier in section \ref{sec:gp}, we consider all the candidate seeds to choose the seed that implies the minimal number of triples to replicate.

The next step extends the partitions with replicates. This is relatively straightforward using Spark's \texttt{union} operator.

Finally,  for querying purpose, each query is extended with a predicate  enforcing local evaluation by joining triples with the same partition identifier.

\subsection{Hybrid approach}
This approach is mixing the subject-based hashing method with the WARP workload-aware processing. 
Hence, using our standard representations of triples and quads together with Spark's ability to easily handle data transformations made our coding effort for this experiment relatively low.

\section{Experimental setting}

\subsection{Datasets and queries}
In this evaluation, we are using one synthetic  and one real world dataset. The synthetic data set corresponds to the well-established LUBM \cite{DBLP:journals/ws/GuoPH05}. 
We are using three instances of LUBM, denoted LUBM1K, LUBM2K and LUBM10K which are parameterized respectively with 1000, 2000 and 10000 universities.
The real world data set consists in Wikidata \cite{DBLP:journals/cacm/VrandecicK14}, a free collaborative knowledge base which will replace Freebase \cite{Bollacker:2008:FCC:1376616.1376746} in 2015.
Table \ref{trip2} presents the number of triples as well as the size of each of these data sets.

\begin{table}
\begin{center}
\begin{tabular}{|c|c|c|}
\hline
Data set  & \#triples & nt File Size\\
\hline
LUBM 1K & 133 M & 22 GB\\
\hline
LUBM 2K & 267 M& 43 GB\\
\hline
LUBM 10K & 1,334 M & 213 GB\\
\hline
Wikidata & 233 M & 37 GB\\
\hline
\end{tabular}
\caption{Dataset statistics of our running examples}\label{trip2}
\end{center}
\end{table}

Concerning queries, we have selected three SPARQL queries from LUBM (namely queries \#2, \#9 and \#12 respectively denoted Q2, Q3 and Q4) and we have created an additional one (denoted Q1) which requires a 3-hop guarantee to be performed
locally on the nHopDB, WARP and hybrid implementations.
To complement the query evaluation, we have created two queries for the Wikidata  experiments, resp. Q5 and Q6. The first one takes the form of a 3-hop property chain query that shows to be much more selective than the LUBM ones, 
the second one is shaped as a simple star and was motivated by the absence of such a form in our query set. All six queries are presented in Appendix \ref{appendix:query}.
\subsection{Computational environment}
Our evaluation was deployed on a cluster consisting of 21 DELL PowerEdge R410 running a Debian distribution with a 3.16.0-4-amd64 kernel version. Each machine has 64GB of DDR3 RAM, two Intel Xeon E5645 processors each of which is equipped with 6 cores running at 2.40GHz and allowing to run two threads in parallel (hyperthreading). Hence, the number of virtual cores amounts to 24 but we used only 15 cores per machine.
In terms of storage, each machine is equipped with a 900GB 7200rpm SATA disk. The machines are connected via a 1GB/s Ethernet Network adapter.
We used Spark version 1.2.1 and implemented all experiments in Scala, using version 2.11.6. 
The Spark setting requires that the total number of cores of the cluster to be specified.
Since in our experiments we considered clusters of  5, 10 and 20 machines respectively, we had to set the number of cores to 75, 150 and 300 cores respectively.

\section{Experimentation}
Since we could not get any query workloads for Wikidata, it was not possible to conduct experimentations with WARP and the hybrid approach over this datasets. 
Moreover, since Metis is limited to datasets of half a million edges, it was not possible to handle nHopDB and WARP over LUBM10K.
Given the fact that the hybrid system relies on subject hashing, and not Metis, it was possible to conduct this experimentation over LUBM10K for that system.

\subsection{Data preparation}
\label{datePrep}

\begin{figure}
\centering
\includegraphics[scale=0.34]{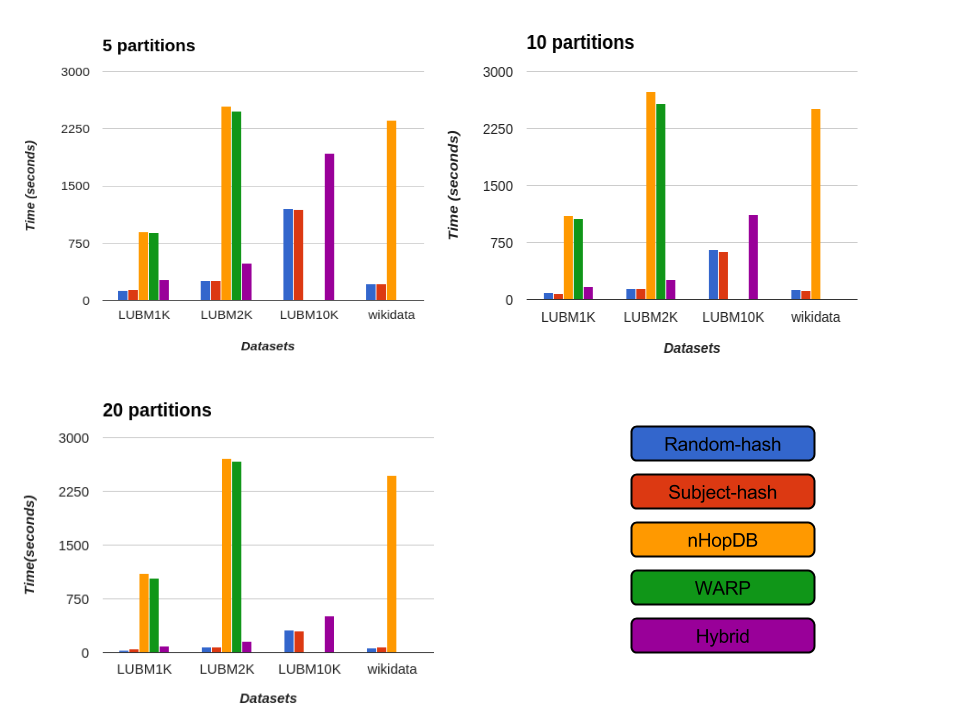}
\caption{Data preparation times}
\label{dataPrep}
\end{figure}
Figure \ref{dataPrep} presents the data preparation processing times for the  different systems. 
As one would expect, the hash-based approaches are much more efficient than the graph partition-based approaches, between 6 and 30 times faster depending on the number partitions.
This is mainly due to the fact that Metis runs on a single machine (we have not tested parMetis, parallelized version of Metis) while the hash operations are being performed in parallel on the Spark cluster.
The evaluation also emphasizes that the hybrid approach presents an interesting compromise between these distribution method families.
By evaluating the different processing steps in each of the solutions, we find out that, for hash-based approaches, around 15\% of  processing time is spent on loading the  datasets whereas the remaining 85\% of time is spent on partitioning the data.
For the graph partitioning approaches, 85 to 90\% corresponds to the time spent by Metis for creating the partitions; the durations increase with the larger dataset sizes.
This explains that the time spent by graph partitioning approaches are slightly increasing even when more machines are added. This does not apply for the other solutions where more machines lead to a reduction of the preparation processing time.


\subsection{Balanced storage}
\begin{figure}
\includegraphics[scale=0.34]{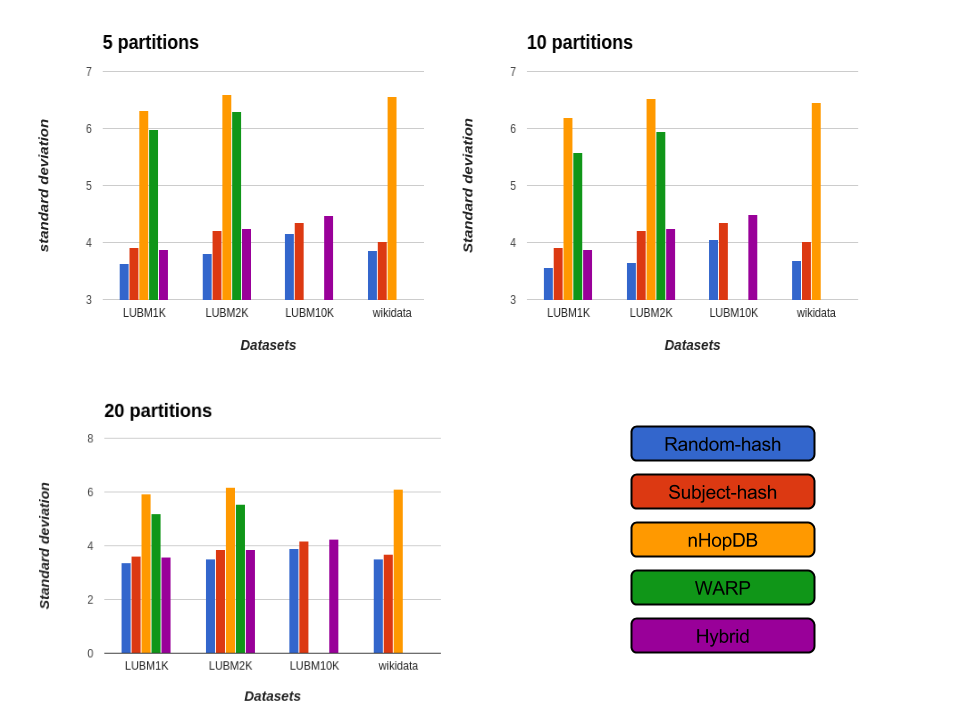}
\caption{Standard deviation}
\label{stddev}
\end{figure}

Load balancing is an important aspect when distributing data for storage and querying purposes. In Figure \ref{stddev}, we present the standard deviations (in log scale) for our different systems. 
For the graph partitioning-based and hybrid approaches, we only consider the standard deviation of the partition sizes at the end of the partitioning process, $i.e.$, Metis partitioning and n-hop guarantee application.

The two hash-based approaches and the hybrid approach are the best solutions and are close to each other. 
This is rather obvious since the hash partitioning approaches are concentrating on load balancing while a graph partitioner tries to reduce the number of edges cut during the fragmentation process.
The hybrid approach is slightly less well-balanced due to the application of the WARP query workload-aware strategy. 
The random-based hashing has 5 to 12\% less deviation  than subject hashing. This is due to high degree nodes that may increase the size  of some partitions.
Although ranging in similar standard deviation values, the nHopDB approach is the less efficient of the graph partitioning solutions. We believe that this is highly related to the number of queries one considers in the query workload.
We consider that further analysis needs to be conducted on real world datasets and query workloads to confirm these nevertheless interesting conclusions.


\subsection{Data replication}
Intrinsically, all solutions present some node replications since a given node can be an object in one partition and a subject in another one. This corresponds to the 1-hop guarantee that ensures validity of data.
In this section, we are only interested in triple replication. Only the nHopDB, WARP and hybrid solutions present such replications. 

Table \ref{trip2} provides the replication rates for each of these systems for the LUBM 1K and 2K datasets. Several conclusions can be drawn from this table. First, Metis-based approaches are more efficient than the subject-hashing of the hybrid system. Remember that by minimizing edge cut, a graph partitioner groups the nodes that are close to each other in the input graph. Secondly, the more partitions the cluster contains, the more overall replication one obtains. The n-hop guarantee replicates less than the query workload-aware method of WARP. Finally, we can stress that the replication of the hybrid approach can be considered quite acceptable given the data replication duration highlighted in Section \ref{datePrep}.
\begin{table}
\begin{center}
\begin{tabular}{|c|c|c|c|c|c|c|c|c|c|}
\hline
 Part. scheme & \multicolumn{3}{|c|}{nHopDB}& \multicolumn{3}{|c|}{WARP} & \multicolumn{3}{|c|}{Hybrid}\\
\hline
Data set  & 5 part.& 10 part. & 20 part.&5 part.& 10 part. & 20 part.& 5 part.& 10 part. & 20 part.\\
\hline
LUBM 1K & 0.12 & 0.16& 0.17& 0.26 & 0.54 & 0.57 & 0.54 & 1.33 & 1.84\\
\hline
LUBM 2K & 0.12 & 0.16 & 0.18& 0.34 & 0.52& 0.54 & 0.54 & 1.33 & 1.94\\
\hline
\end{tabular}
\caption{Replication rate comparison for three partitioning schemes and three cluster sizes}\label{trip2}
\end{center}
\end{table}

\subsection{Query processing}
In order to efficiently process local queries and to fairly support performance comparison in a distributed setting, we must use the same computing resources for local and distributed runs. 
A local query runs in parallel when every machine only has to access its own partition. 
To exploit the multicore machines on which we perform our experiments, it is interesting to consider not only inter-partition parallelism but intra-partition parallelism as well.
Unfortunately, intra-partition parallelism is not fully supported in Spark since a partition is the unit of data that one core is processing. Thus, to use 15 cores on a machine, we must split a partition into 15 sub partitions.
Spark does not allow to specify that such sub-partitions must reside together on the same machine. We expect that future version of Spark will allow such control.
In the absence of any triple replication, the hash-based solutions are not impacted by this limitation. This is not the case for the systems using replication. 
For instance, for the two query workload-aware solutions (i.e., WARP and hybrid), we conducted our experiment using a workaround that forces Spark to use only one machine for one partition: for local queries, we run Spark with only one slave machine.
Then we load only the data of one partition and process the query locally in parallel using all the cores. 
To be fair and take into account the possibility that a local query might run faster in some partitions than in some other partitions, we repeat the experiment for every partition and report the maximum response time.

The case of nHopDB is more involved and requires to develop a special dedicated query processor, specialized for Spark, to fully benefit from the data fragmentation. 
In a nutshell, that system would have to combine intra and inter-partition query processor. The former would run for query subgraphs that can run locally and the second one would perform joins over all partitions with retrieved temporary results. 
Since the topic of this paper concerns the evaluation of distribution strategies, we do not detail the implementation of such a query processor in this work and hence we do not present any results for the nHopDB system.

\begin{figure}
\centering
\includegraphics[scale=0.34]{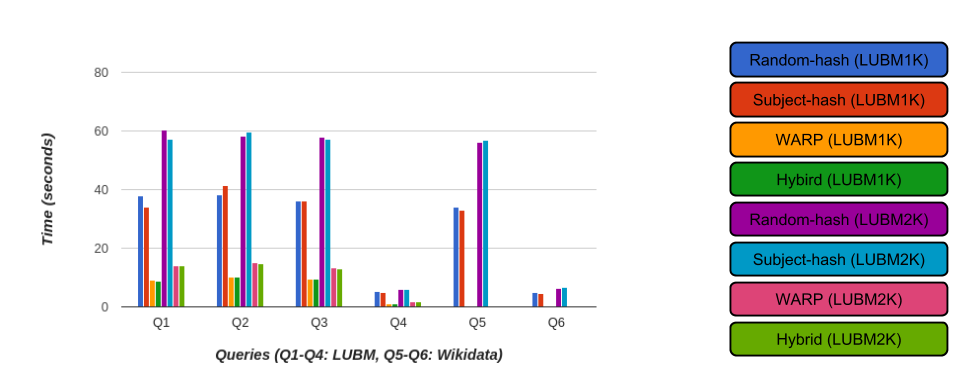}
\caption{Query Evaluation on 20 partitions}
\label{queryEval}
\end{figure}

Table \ref{queryEval} presents the query processing times for our dataset. Due to space limitation, we only present the execution time obtained over the 20 partitions experiment. 
The web site companion (see \cite{siteweb}) highlights that the more partitions the more efficient the query processing. 
The table clearly highlights that the WARP systems are more efficient than the hashing based solutions. Obviously, the simpler the query, $e.g.$, Q4 and Q6, run locally while the others require inter-partition communication. 
With the Spark version ($i.e.$, 1.2.1) we were conducting this experiment on, we could not measure the inter node information communication. In fact, Spark's \texttt{shuffle read} measure indicates the total information exchange
(locally on a node and globally over the network).

\section{Related work}
Some other interesting works have recently been published on the distribution of RDF data. Systems such as Semstore \cite{Wu:2014:SSD:2661829.2661876} and SHAPE \cite{DBLP:journals/pvldb/LeeL13} take some original position. 
Instead of using the common query workload, Semstore divides a complete RDF graph into a set of paths which cover all the original graph nodes, possibly with node overlapping between paths. These paths are denoted Rooted Sub Graph (RSG in short) since they are generated starting from nodes with a null in-degree, $i.e.$, roots, to all their possible leaves. A special workaround is used to handle cycles that may occur at the root position, $i.e.$, cycles that are not reachable from any root. The idea is then to regroup these RSG into different partitions. This is obviously a hard problem for which the authors propose an approximated solution. Their solution uses the K-means clustering approach which regroups RSG with common segments together in the same partition. A first limitation of this approach is the high dimensionality of the vectors handled by the K-means algorithm, $i.e.$, the size of any vector corresponds to the number of nodes in the graph. A second limitation is related to the lack of an efficient balancing of the number triples across the partitions. In fact, the system operates at the coarse-grained level of RSG and provides a balancing at this level only. Semstore is finally limited in terms of join patterns. It can efficiently handle S-O (subject-object) and  S-S (subject-subject) join patterns but other patterns, such as O-O (object-object) may require inter node communication. 

The motivation of the SHAPE system is that graph partitioning approaches do not scale. Just like in our hybrid solution, they propose to replace the graph partitioning step by a hash partitioning one. 
Then, just like in the nHopDB system, they replicate according to the n-hop guarantee. Hence, they do not consider any query workload and take the risk of inter-partition communication for long chain queries longer than their n-hop guarantee.

\section{Conclusions and perspectives}
This paper presents an evaluation of distributed systems ranging over two important partitioning categories: hashing and graph partitioning. 
The choice of using the Spark framework is motivated by its high performance. For certain operations, it is considered to be 100 times faster than Hadoop MapReduce.
While several systems have been designed on top of Hadoop, we are not aware of any RDF data management systems running on top of Spark.
The main motivation of the experiments is that existing partitioning solutions do not scale gracefully to several billion triples.
For instance, the Metis partitioner is limited to less than half a billion triples and SemStore (cf. related works section) relies on K-Means clustering of vectors whose dimension amount to the number of nodes of the data to be processed 
($i.e.$, 32 millions in the case of LUBM1K). Computing a distance at such high dimension is currently not possible within Spark, even when using sparse vectors. Moreover, applying a dimension reduction algorithm to all the vectors is not tractable.

The conclusion of our experiment is that basic hash-based partitioning solutions are viable for distributed RDF management:
 they come at no preparation cost, $i.e.$, it only requires to load the triples into the right machine, and it is fully supported by the underlying Spark system. 
 As emphasized by our experimentation, Spark scales out to several billion triples by simply adding extra machines. Nevertheless, without any replication, these systems may hinder availability and reduce the parallelism of query processing. 
 They also involve a lot of network communications for complex queries which require to retrieve data from many partitions. Nonetheless, by making intensive use of main memory, we believe that Spark provides a high potential for these systems.
 Clearly, with the measures we have obtained in this evaluation, we can stress that if one needs a fast access to large RDF datasets and is, to some extent, ready to sacrifice the performance of processing complex query patterns 
 then the hash-based solution over Spark is a good compromise.
  
Concerning the nHopDB and WARP approaches, we consider that using Metis is an important drawback. Based in these observations, we investigated the hybrid  candidate solution which does not involve a heavy preparation step and 
retains the interesting query workload aware replication strategy. This approach may be particularly interesting for data warehouses where the most common queries (materialized views) are well identified. 
With this hybrid solution we may get the best of worlds, the experiments clearly emphasize that the replication overhead compared to the pure WARP approach is marginal but the gain in data preparation is quite important. 

Concerning Spark, we highlighted that it can process distributed RDF queries efficiently. Moreover, the system can be used for the two steps : data preparation and query processing in an homogeneous way. 
Rewriting SPARQL queries into the Scala language (supported by Spark) is rather easy and we consider that there is room for optimization. The next versions of Spark which are supposed to provide more feedback on data exchange over 
the network should help fine-tune our experiments and design a complete production-ready system.
\bibliographystyle{abbrv}
\bibliography{amw15} 

\begin{thebibliography}{10}

\bibitem{siteweb}
\url{http://webia.lip6.fr/{$\sim$}baazizi/research/iswc2015eval/expe.html}.

\bibitem{Bollacker:2008:FCC:1376616.1376746}
K.~Bollacker, C.~Evans, P.~Paritosh, T.~Sturge, and J.~Taylor.
\newblock Freebase: A collaboratively created graph database for structuring
  human knowledge.
\newblock In {\em Proceedings of the 2008 ACM SIGMOD International Conference
  on Management of Data}, SIGMOD '08, pages 1247--1250, New York, NY, USA,
  2008. ACM.

\bibitem{cai2004rdfpeers}
M.~Cai and M.~Frank.
\newblock {RDFPeers}: A scalable distributed {RDF} repository based on a
  structured peer-to-peer network.
\newblock In {\em Proc.\ 13th International World Wide Web Conference}, New
  York City, NY, USA, May 2004.

\bibitem{cure:hal-01083133}
O.~Cur{\'e} and G.~Blin.
\newblock {\em {RDF Database Systems, 1st Edition}}.
\newblock {Morgan Kaufmann}, Nov. 2014.

\bibitem{DBLP:conf/osdi/DeanG04}
J.~Dean and S.~Ghemawat.
\newblock Mapreduce: Simplified data processing on large clusters.
\newblock In {\em OSDI}, pages 137--150, 2004.

\bibitem{DBLP:journals/debu/Erling12}
O.~Erling.
\newblock Virtuoso, a hybrid rdbms/graph column store.
\newblock {\em IEEE Data Eng. Bull.}, 35(1):3--8, 2012.

\bibitem{DBLP:conf/dac/FiducciaM82}
C.~M. Fiduccia and R.~M. Mattheyses.
\newblock A linear-time heuristic for improving network partitions.
\newblock In {\em Proceedings of the 19th Design Automation Conference, {DAC}
  '82, Las Vegas, Nevada, USA, June 14-16, 1982}, pages 175--181, 1982.

\bibitem{DBLP:journals/corr/abs-1212-5636}
L.~Galarraga, K.~Hose, and R.~Schenkel.
\newblock Partout: {A} distributed engine for efficient {RDF} processing.
\newblock {\em CoRR}, abs/1212.5636, 2012.

\bibitem{DBLP:journals/ws/GuoPH05}
Y.~Guo, Z.~Pan, and J.~Heflin.
\newblock Lubm: A benchmark for owl knowledge base systems.
\newblock {\em J. Web Sem.}, 3(2-3):158--182, 2005.

\bibitem{DBLP:conf/sigmod/GurajadaSMT14}
S.~Gurajada, S.~Seufert, I.~Miliaraki, and M.~Theobald.
\newblock Triad: a distributed shared-nothing {RDF} engine based on
  asynchronous message passing.
\newblock In {\em International Conference on Management of Data, {SIGMOD}
  2014, USA, June 22-27, 2014}, pages 289--300, 2014.

\bibitem{DBLP:journals/pvldb/HammoudRNBS15}
M.~Hammoud, D.~A. Rabbou, R.~Nouri, S.~Beheshti, and S.~Sakr.
\newblock {DREAM:} distributed {RDF} engine with adaptive query planner and
  minimal communication.
\newblock {\em {PVLDB}}, 8.

\bibitem{DBLP:conf/semweb/HarthUHD07}
A.~Harth, J.~Umbrich, A.~Hogan, and S.~Decker.
\newblock {YARS2:} {A} federated repository for querying graph structured data
  from the web.
\newblock In {\em The Semantic Web, 6th International Semantic Web Conference,
  {ISWC} 2007 + {ASWC} 2007, Busan, Korea, November 11-15, 2007.}, pages
  211--224, 2007.

\bibitem{DBLP:conf/icde/HoseS13}
K.~Hose and R.~Schenkel.
\newblock {WARP:} workload-aware replication and partitioning for {RDF}.
\newblock In {\em Workshops Proceedings of the 29th {IEEE} International
  Conference on Data Engineering, {ICDE} 2013}, pages 1--6, 2013.

\bibitem{DBLP:journals/pvldb/HuangAR11}
J.~Huang, D.~J. Abadi, and K.~Ren.
\newblock Scalable sparql querying of large rdf graphs.
\newblock {\em PVLDB}, 4(11):1123--1134, 2011.

\bibitem{Karypis:1998:FHQ:305219.305248}
G.~Karypis and V.~Kumar.
\newblock A fast and high quality multilevel scheme for partitioning irregular
  graphs.
\newblock {\em SIAM J. Sci. Comput.}, 20(1):359--392, Dec. 1998.

\bibitem{DBLP:journals/pvldb/LeeL13}
K.~Lee and L.~Liu.
\newblock Scaling queries over big {RDF} graphs with semantic hash
  partitioning.
\newblock {\em {PVLDB}}, 6(14):1894--1905, 2013.

\bibitem{DBLP:conf/www/NejdlWQDSNNPR02}
W.~Nejdl, B.~Wolf, C.~Qu, S.~Decker, M.~Sintek, A.~Naeve, M.~Nilsson,
  M.~Palm{\'{e}}r, and T.~Risch.
\newblock {EDUTELLA:} a {P2P} networking infrastructure based on {RDF}.
\newblock In {\em Proceedings of the Eleventh International World Wide Web
  Conference, {WWW} 2002, USA}, pages 604--615, 2002.

\bibitem{DBLP:journals/vldb/NeumannW10}
T.~Neumann and G.~Weikum.
\newblock The rdf-3x engine for scalable management of rdf data.
\newblock {\em VLDB J.}, 19(1):91--113, 2010.

\bibitem{Rohloff:2010:HMS:1940747.1940751}
K.~Rohloff and R.~E. Schantz.
\newblock High-performance, massively scalable distributed systems using the
  mapreduce software framework: The shard triple-store.
\newblock In {\em Programming Support Innovations for Emerging Distributed
  Applications}, PSI EtA '10, pages 4:1--4:5, New York, NY, USA, 2010. ACM.

\bibitem{Sadalage:2012:NDB:2381014}
P.~J. Sadalage and M.~Fowler.
\newblock {\em NoSQL Distilled: A Brief Guide to the Emerging World of Polyglot
  Persistence}.
\newblock Addison-Wesley Professional, 1st edition, 2012.

\bibitem{DBLP:journals/cacm/StonebrakerADMPPR10}
M.~Stonebraker, D.~J. Abadi, D.~J. DeWitt, S.~Madden, E.~Paulson, A.~Pavlo, and
  A.~Rasin.
\newblock Mapreduce and parallel dbmss: friends or foes?
\newblock {\em Commun. {ACM}}, 53(1):64--71, 2010.

\bibitem{DBLP:journals/cacm/VrandecicK14}
D.~Vrandecic and M.~Kr{\"{o}}tzsch.
\newblock Wikidata: a free collaborative knowledgebase.
\newblock {\em Commun. {ACM}}, 57(10):78--85, 2014.

\bibitem{Wu:2014:SSD:2661829.2661876}
B.~Wu, Y.~Zhou, P.~Yuan, H.~Jin, and L.~Liu.
\newblock Semstore: A semantic-preserving distributed rdf triple store.
\newblock In {\em Proceedings of the 23rd ACM International Conference on
  Conference on Information and Knowledge Management}, CIKM '14, pages
  509--518, New York, NY, USA, 2014. ACM.

\bibitem{DBLP:conf/nsdi/ZahariaCDDMMFSS12}
M.~Zaharia, M.~Chowdhury, T.~Das, A.~Dave, J.~Ma, M.~McCauly, M.~J. Franklin,
  S.~Shenker, and I.~Stoica.
\newblock Resilient distributed datasets: {A} fault-tolerant abstraction for
  in-memory cluster computing.
\newblock In {\em Proceedings of the 9th {USENIX} Symposium on Networked
  Systems Design and Implementation, {NSDI} 2012, San Jose, CA, USA, April
  25-27, 2012}, pages 15--28, 2012.

\bibitem{DBLP:conf/hotcloud/ZahariaCFSS10}
M.~Zaharia, M.~Chowdhury, M.~J. Franklin, S.~Shenker, and I.~Stoica.
\newblock Spark: Cluster computing with working sets.
\newblock In {\em 2nd {USENIX} Workshop on Hot Topics in Cloud Computing,
  HotCloud'10, Boston, MA, USA, June 22, 2010}, 2010.

\bibitem{DBLP:journals/pvldb/ZengYWSW13}
K.~Zeng, J.~Yang, H.~Wang, B.~Shao, and Z.~Wang.
\newblock A distributed graph engine for web scale {RDF} data.
\newblock {\em {PVLDB}}, 6(4):265--276, 2013.

\end{thebibliography}

\appendix
\section{Queries}We now present the six queries that have been used during our evaluation.
\label{appendix:query}

\noindent {\bf Q1}: \texttt{SELECT ?x ?y ?z WHERE \{?x lubm:advisor ?y. ?y lubm:worksFor ?z.\\ ?z lubm:subOrganisation ?t.\}}

\noindent {\bf Q2}: \texttt{SELECT ?x ?y ?z WHERE \{?x rdf:type lubm:GraduateStudent.\\ ?y rdf:type lubm:University.  ?z rdf:type lubm:Department.\\ ?x lubm:memberOf ?z. ?x lubm:subOrganizationOf ?y. \\
  ?x lubm:undergraduateDegreeFrom ?y\}}

\noindent {\bf Q3}: \texttt{SELECT ?x ?y ?z WHERE \{{?x rdf:type lubm:Student.\\ ?y rdf:type lubm:Faculty.?z rdf:type lubm:Course.?x lubm:advisor ?y. ?y lubm:teacherOf ?z. ?x lubm:takesCourse ?z}\}}

\noindent {\bf Q4}: \texttt{SELECT ?x ?y WHERE \{?x rdf:type lubm:Chair.\\ ?y rdf:type lubm:Department. ?x lubm:worksFor ?y.\\?y lubm:subOrganizationOf <http://www.University0.edu>\}}

\noindent {\bf Q5}: \texttt{SELECT ?x ?y ?z WHERE \{?x entity:P131s ?y. ?y entity:P961v> ?z. ?z entity:P704s ?w.\}}

\noindent {\bf Q6}: \texttt{SELECT ?x ?y ?z WHERE  \{?x entity:P39v ?y. ?x entity:P580q ?z. ?x rdf:type ?w\}}
\end{document}